\documentclass[conference]{IEEEtran}
\IEEEoverridecommandlockouts

\usepackage{caption}

\usepackage{eucal}
\usepackage{placeins}
\usepackage{makecell}
\usepackage{svg}
\usepackage{array}
\usepackage{tabularx}
\usepackage{booktabs}
\usepackage{multirow}
\usepackage{cite}
\usepackage{amsmath,amssymb,amsfonts}
\usepackage{algorithm}
\usepackage{algpseudocode}
\usepackage{graphicx}
\usepackage{textcomp}
\usepackage{xcolor}
\usepackage{graphicx}
\usepackage{subcaption}
\def\BibTeX{{\rm B\kern-.05em{\sc i\kern-.025em b}\kern-.08em
    T\kern-.1667em\lower.7ex\hbox{E}\kern-.125emX}}

\usepackage{fancyhdr}
\usepackage{hyperref}

\fancypagestyle{firstpage}{
    \fancyhf{}

    \fancyfoot[C]{%
        \footnotesize
        \href{khaled.jebari@imt-atlantique.net}{khaled.jebari@imt-atlantique.net}
        \;\textbar\;
        \href{luiz.anetneto@imt-atlantique.fr}{luiz.anetneto@imt-atlantique.fr}
        \;\textbar\;
        \href{pyndiah.ramesh@wanadoo.fr}{pyndiah.ramesh@wanadoo.fr}
        \;\textbar\;
        \href{jl.debougrenet@imt-atlantique.fr}{jl.debougrenet@imt-atlantique.fr}
    }
}
\begin{document}

\title{Multidimensional Quantum Block Turbo Codes}

\author{
    \IEEEauthorblockN{
        Khaled Jebari\IEEEauthorrefmark{1},
        Luiz Anet Neto\IEEEauthorrefmark{1}\textsuperscript{,}\IEEEauthorrefmark{2},
        Ramesh Pyndiah\IEEEauthorrefmark{3},
        Jean-Louis de Bougrenet de la Tocnaye\IEEEauthorrefmark{1}
    }
    \IEEEauthorblockA{
        \IEEEauthorrefmark{1}\textit{Optics Department}, \textit{IMT Atlantique}, 655 Av. du Technopôle, 29280 Plouzané, France. \\
        \IEEEauthorrefmark{2}\textit{Lab-STICC CNRS UMR 6285}, Technopôle Brest-Iroise - CS 83818, 29238 Brest Cedex 3, France. \\
        \IEEEauthorrefmark{3}\textit{Previously with the Optics Department}, \textit{IMT Atlantique}, 655 Av. du Technopôle, 29280 Plouzané, France.
    }

}

\maketitle
\thispagestyle{firstpage}

\begin{abstract}
The theory of turbo iterative decoding was brought into the quantum coding framework through the development of quantum serial turbo codes and quantum block turbo codes. In this paper, we present a theory for multidimensional quantum block turbo codes, a turbo decoding technique for multi layer product codes. We describe its algorithm and simulate the performance for the 3D case on a depolarizing
channel.  
\end{abstract}


\section{Introduction}

Turbo iterative decoding provides an efficient framework for approximating maximum-likelihood decoding through the iterative exchange of soft information between two constituent decoding blocks. This approach has been successfully extended to quantum error-correcting codes, including quantum serial turbo codes (QSTC) \cite{poulin2009quantumserialturbocodes} and quantum block turbo codes (QBTC) \cite{Jebari2026}.

The construction of QBTCs combines two constituent stabilizer codes through successive row and column encoding operations, resulting in a product-like structure. The associated iterative decoding process relies on the exchange of soft information between two soft-input soft-output (SISO) decoders to progressively refine error estimates. This exchange involves marginalization and reorganization procedures, enabling each constituent decoder to improve its estimates using information provided by the other. By exploiting syndrome information and accounting for quantum degeneracy, QBTCs provide a promising framework for iterative quantum error correction.

We study here the extendability of the QBTC theory to multidimensional product codes as anticipated in \cite{Jebari2026}. We introduce their decoding algorithm and then we apply the concept to the 3D case and we examine their turbo performance under a depolarizing channel. The present work naturally builds upon \cite{Jebari2026}, to which we refer the reader for further details.

This paper is organized as follows. We start with some
preliminaries on multidimensional product codes in Section.\ref{sec:multi product}. We then provide in Section.\ref{sec:multi}
an overview of their QBTC decoding structure. In Section.\ref{sec:sim}, we present some simulation results for the 3D case and we conclude our work in Section.\ref{sec:concl}.

\begin{table}[b]
\centering
\caption{Combinations of QBTCs.}
\resizebox{\columnwidth}{!}{%
\begin{tabular}{c c c c c}
\hline
code & code rate & $k$ & $n$ & distance   \\
\hline
QBTC $\left[\!\left[5,1,3\right]\!\right] \times \left[\!\left[5,1,3\right]\!\right]\times \left[\!\left[5,1,3\right]\!\right]$
 & $0,008$ & $1$ & $125$ & $27$  \\
QBTC $\left[\!\left[4,2,2\right]\!\right] \times \left[\!\left[4,2,2\right]\!\right]\times \left[\!\left[4,2,2\right]\!\right]$
 & $0,125$ & $8$ & $64$ & $8$  \\
QBTC $\left[\!\left[8,3,3\right]\!\right] \times \left[\!\left[8,3,3\right]\!\right]\times \left[\!\left[8,3,3\right]\!\right]$
 & $0,052$ & $27$ & $512$ & $27$ \\
QBTC $\left[\!\left[11,5,3\right]\!\right] \times \left[\!\left[11,5,3\right]\!\right]\times \left[\!\left[11,5,3\right]\!\right]$
 & $0,093$ & $125$ & $1331$ &  $27$ \\
QBTC $\left[\!\left[15,7,3\right]\!\right] \times \left[\!\left[15,7,3\right]\!\right] \times \left[\!\left[15,7,3\right]\!\right]$
 & $0,101$ & $343$ & $3375$ & $27$\\
 QBTC $\left[\!\left[8,3,3\right]\!\right] \times \left[\!\left[11,5,3\right]\!\right] \times \left[\!\left[15,7,3\right]\!\right]$
 & $0,079$ & $105$ & $1320$ & $27$ \\

\hline
\end{tabular}%
}

\vspace{2mm}

\label{tab:comb}
\end{table}

\section{Multidimensional product codes} \label{sec:multi product}

Multilayer concatenation offers a scalable approach to quantum error correction by spreading redundancy across several encoding levels, improving the protection of quantum information while maintaining a balance between error correction performance, qubit overhead, and coding rate. Concatenated codes have received renewed interest, driven by advances in low-overhead fault-tolerant protocols as studied in \cite{Yamasaki_2024,Yoshida_2025,goto2024manyhypercube,xbzn-vn37,goto2026optimizedmanyhypercubecodeslower}.

In a multilayer concatenation scheme, the overall coding rate after \(L\) layers is given by \(R^{(L)}=\prod_{l=1}^{L}R_l\), where \(R_l\) denotes the rate introduced at the \(l\)-th layer. If a fixed-rate code with \(R_l=R<1\) is repeatedly concatenated, the resulting rate decreases exponentially as \(R^L\) and therefore vanishes in the large-\(L\) limit. In contrast, with clever choice of constituent codes, we could mitigate the vanishing problem. For example, the concatenated Hamming-code construction avoids this vanishing-rate behavior by using component codes of increasing block size whose individual rates progressively approach unity \cite{Yamasaki_2024}. For quantum Hamming codes $\mathcal{Q}_{r_l}=\left[\!\left[2^{r_l}-1,2^{r_l}-2r_l-1,3\right]\!\right]$ with $r_l=l+2$, \(R_l=K_{r_l}/N_{r_l}=1-2r_l/(2^{r_l}-1)\), so that, for an appropriate growth of \(r_l\), the accumulated rate loss remains finite and the infinite product \(\prod_l R_l\) converges to a strictly positive value denoted $\frac{1}{\eta_{\infty}}$. It has been demonstrated numerically in \cite{Yamasaki_2024} that $\eta_{\infty}<36$ and saturates at around $8$ levels. Thus, the rate is $\geq \frac{1}{\eta_{\infty}}>\frac{1}{36}\approx 0.027>0$. Consequently, additional concatenation layers can provide increasing error protection without causing the asymptotic coding rate to collapse to zero. 

In addition, this factor could be made close to $1$ by starting the concatenations from $Q_{r_8}$ rather than $Q_3$ \cite{Yamasaki_2024}. More generally, concatenated quantum Bose-Chaudhuri-Hocquenghem (BCH) \cite{PhysRevA.54.4741} codes $ \bigl[\!\bigl[2^{r_l}-1,\;2^{r_l}-1-2tr_l,\;2t+1\bigr]\!\bigr], t\ge1,
$ can retain a non-vanishing rate while achieving larger distances than quantum Hamming codes, potentially improving error suppression as \(L\) increases \cite{Yamasaki_2024}.

\section{Multidimensional QBTC} \label{sec:multi}

\begin{figure*}[t]
\vspace{-0.4cm}
\centerline{\includegraphics[width=\linewidth]{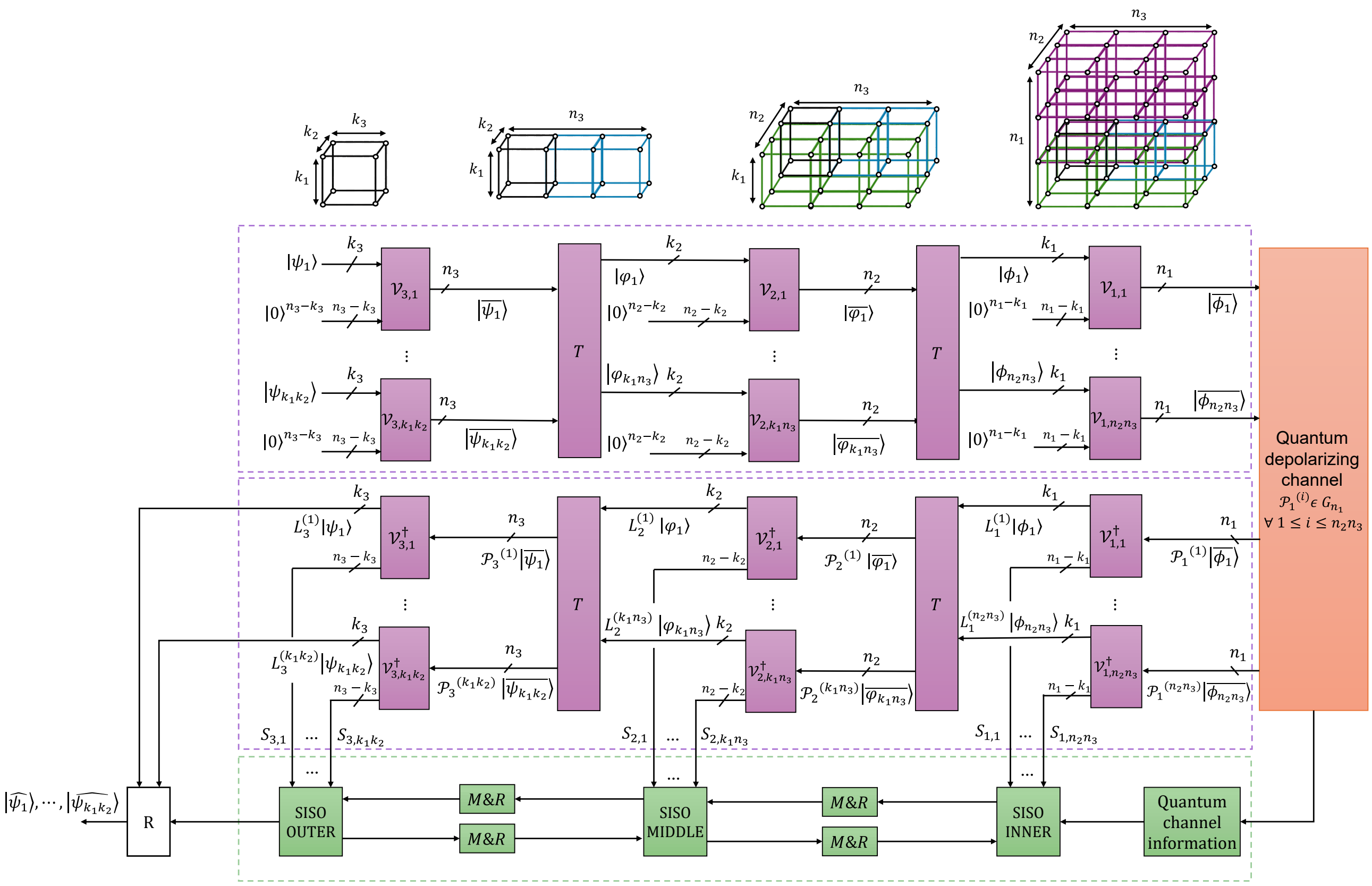}}
\caption{3D QBTC encoding and decoding: product code $(n_1, k_1) \times (n_2, k_2) \times (n_3, k_3)$. In the cubic representation shown at the top, the qubits are represented by vertices. The SISO and M \& R blocks broadly follow the algorithms presented in \cite{Jebari2026}.}
\label{fig:multi layer}
\end{figure*}

The QBTC framework can be naturally extended to the encoding and decoding of multidimensional product codes by adopting a multilayer architecture, as illustrated in Fig.\ref{fig:multi layer}. The SISO units remain identical to those used in the two-dimensional case studied in \cite{Jebari2026,jebari2026highratequantumblock}. The main modification consists in introducing an additional layer of SISO blocks for each intermediate layer between the inner and outer decoders located at the two extremities of the decoding chain.

These SISO blocks iteratively exchange soft information through marginalization and reorganization (M \& R) operations \cite{Jebari2026}, propagating information from the inner layer toward the outer layer and subsequently in the reverse direction. This bidirectional exchange enables the iterative decoding process to exploit correlations across all layers of the product-code structure. We would like to emphasize that the marginalization operation could be directly incorporated into the SISO algorithm. Consequently, both the inputs and outputs will be expressed in terms of marginalized probabilities, while the inter-layer connections perform the necessary reorganizations. This formulation provides greater flexibility and enables the freedom of choice in connectivity between the different layers like in the case of quantum holographic turbo codes \cite{JebariQHTC2026}.

Increasing the number of layers may further strengthen the turbo effect by enabling more extensive exchange of extrinsic information between constituent decoders, while simultaneously allowing the construction of product codes with larger minimum distances. This will be investigated in the next section.

Focusing on the 3D QBTC case and using constituent codes $C_1=(n_1,k_1),C_2=(n_2,k_2)$ and $C_3=(n_3,k_3)$ respectively as inner, middle and outer codes, we obtain $k_1k_2$ outer SISO blocks, $k_1n_3$ middle SISO blocks and $n_2n_3$ inner SISO blocks. The computational complexity of the SISO blocks is governed by the relatively small size of the constituent codes \cite{jebari2026highratequantumblock}. Consequently, the use of small constituent blocks remains computationally feasible even when the resulting multilayer product code is substantially larger. This approach enables the construction of higher-dimensional codes from simpler constituent codes, thereby avoiding the exponential increase in decoding complexity associated with larger constituent codes.
In Tab.\ref{tab:comb}, we collect some 3D QBTC combinations for which we examine their turbo performances in the next section. 
 
\section{Simulations and Results} \label{sec:sim}

\begin{figure}[!t]
    \centering
    \includegraphics[width=\linewidth]{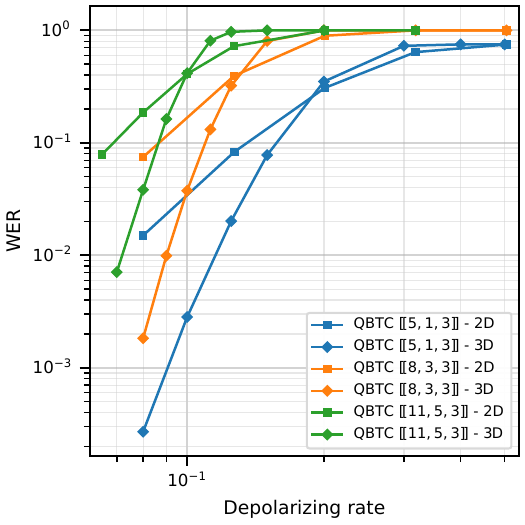}
    \caption{Comparaison of QBTC 2D-3D decoding performance of different code combinations from Table.\ref{tab:comb} after five iterations.}
    \label{fig:every}
\end{figure}

\begin{figure*}[!t]
    \centering

    \begin{subfigure}{0.49\textwidth}
        \centering
        \includegraphics[width=\linewidth]{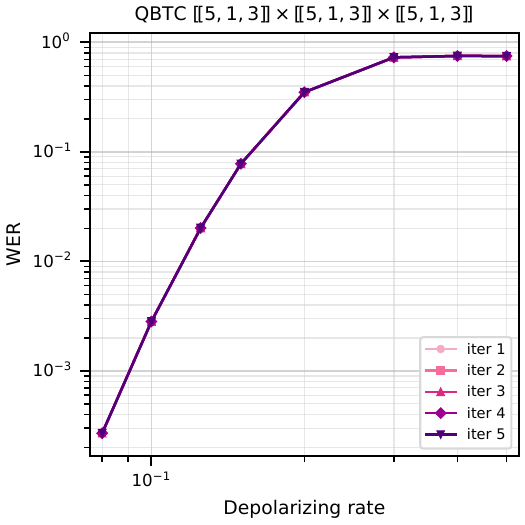}
    \end{subfigure}
    \hfill
    \begin{subfigure}{0.49\textwidth}
        \centering
        \includegraphics[width=\linewidth]{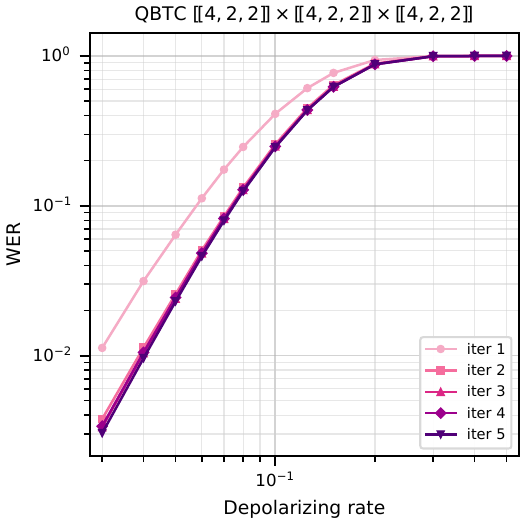}
    \end{subfigure}

    \vspace{0.5cm}

    \begin{subfigure}{0.49\textwidth}
        \centering
        \includegraphics[width=\linewidth]{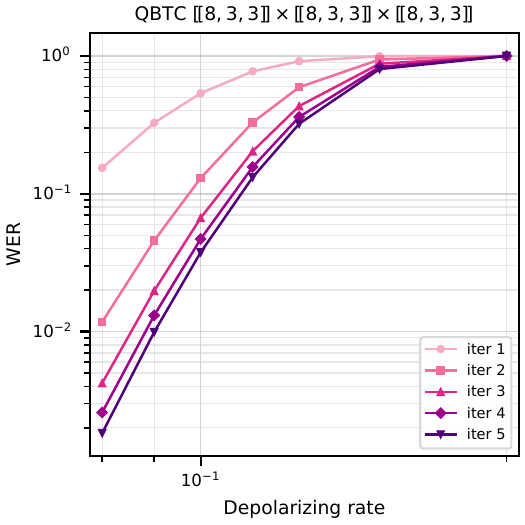}
    \end{subfigure}
    \hfill
    \begin{subfigure}{0.49\textwidth}
        \centering
        \includegraphics[width=\linewidth]{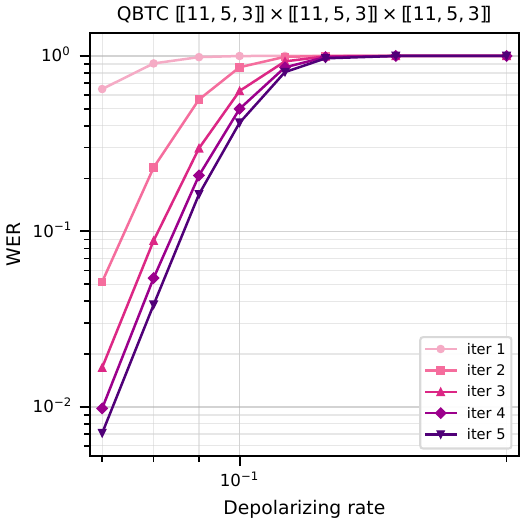}
    \end{subfigure}

    \caption{Performance of 3D QBTCs with variable code parameters.}
    \label{fig:3D QBTC}
\end{figure*}

We present in Fig.\ref{fig:every} and Fig.\ref{fig:3D QBTC} some 2D and 3D QBTCs performances on a depolarizing channel. We observe as expected that QBTC $\left[\!\left[5,1,3\right]\!\right] \times \left[\!\left[5,1,3\right]\!\right] \times \left[\!\left[5,1,3\right]\!\right]$ has no signs of turbo effect as its constituent code only encodes one logical qubit. For the other combinations that use constituent codes with more than one logical qubit, we clearly observe the turbo advantage leading to progressively greater performance improvements over successive iterations, which demonstrates our initial assumption that the turbo effect is more prominent for multidimensional QBTCs \cite{Jebari2026}.
As shown in Fig.\ref{fig:every}, we get after five iterations a threshold around $0.19$ for the QBTC $\left[\!\left[5,1,3\right]\!\right]$, a threshold around $0.13$ for the QBTC $\left[\!\left[8,3,3\right]\!\right]$ and a threshold of $0.1$ for the QBTC $\left[\!\left[11,5,3\right]\!\right]$. This demonstrates the beneficial effect of the 3D turbo decoding, resulting in relatively high threshold values which correspond to the maximum physical error rate below which increasing the code distance effectively reduces logical errors.

\section{Conclusions} \label{sec:concl}

This paper present an iterative decoding technique for multi layer product codes that could offer, with clever construction, an effective balance among code rate, minimum distance, and implementation complexity. This theory is developed as an extension of quantum block turbo codes for multidimensional product codes. We present their decoding algorithm and examine their turbo performance. Consequently, the turbo advantage is more prominent for multidimensional QBTC leading to relatively high thresholds.

\section{LLM Usage Disclaimer}
Large Language Models (LLMs) were used in this work as a coding assistance tool to help optimize our simulation code, with the goal of improving execution speed and enabling efficient parallelization across both Central Processing Unit (CPU) and Graphics Processing Unit (GPU). All generated suggestions were reviewed, adapted, and validated by the authors.

\bibliographystyle{IEEEtran}
\bibliography{references}

\end{document}